\documentclass[%
 reprint,
superscriptaddress,
 amsmath,amssymb,
 aps,
pra,
]{revtex4-2}

\newcommand{\expect}[1]{\left\langle #1 \right\rangle}
\newcommand{\ket}[1]{\left| #1 \right\rangle}

\newcommand{\BerryV}{\mathbf{\tilde{B}}}
\newcommand{\Berry}{\tilde{B}}

\usepackage{graphicx}
\usepackage{dcolumn}
\usepackage{bm}

\begin{document}

\preprint{APS/123-QED}

\title{Tuning Topological Charge and Gauge Field Anisotropy\\ in a Spin-1 Synthetic Monopole}

\author{Nicholas Milson}
\altaffiliation{These authors contributed equally.}
\email{nmilson@ualberta.ca}
\affiliation{%
Department of Physics, University of Alberta, Edmonton T6G 2E1, Canada.
}%

\author{Arina Tashchilina}
\altaffiliation{These authors contributed equally.}
\affiliation{%
Department of Physics, University of Alberta, Edmonton T6G 2E1, Canada.
}%
\affiliation{
 Institut für Quantenoptik und Quanteninformation, Österreichische Akademie der Wissenschaften, Innsbruck 6020, Austria 
}%

\author{Kathleen Tamura}%
\affiliation{%
Department of Physics, University of Alberta, Edmonton T6G 2E1, Canada.
}%

\author{Douglas Florizone}%
\affiliation{%
Department of Physics, University of Alberta, Edmonton T6G 2E1, Canada.
}%

\author{Lindsay~J.~LeBlanc}%
\email{lindsay.leblanc@ualberta.ca}
\affiliation{%
Department of Physics, University of Alberta, Edmonton T6G 2E1, Canada.
}%

\date{\today}

\begin{abstract}
Higher-dimensional Hilbert spaces in quantum simulation, as in all quantum science, expand the range of accessible phenomena.  In this work, we experimentally realize a synthetic monopole using an ultracold spin-1 ensemble, where the monopole charge is quantified by the topologically invariant first Chern number and sources a synthetic magnetic field quantified by the Berry curvature.  By using a three-level system with tunable spin-tensor coupling, we introduce anisotropy to the field, directly measure the Chern number, and observe a topological phase transition.  We verify the robustness of the monopole’s topological charge under deformation, and observe signatures of the topological phases using spin-texture and Majorana-star measurements. This work demonstrates spin-tensor coupling as a tuning parameter for engineering both geometric anisotropy and a rich topological phase space.
\end{abstract}

\maketitle

\noindent
Gauge field theories lie at the foundation of our deepest understanding of physics~\cite{hooft2007making}. While important breakthroughs confirming these theories are made through large-scale physics experiments~\cite{kharzeev2021chiral,acharya2022search}, synthetic quantum matter provides opportunities to engineer and probe artificial gauge fields, including configurations that may not have known analogs in nature, but are governed by the same universal physics. 
Control and measurement of the geometric phase in these systems~\cite{dalibard2011colloquium,goldman2014light, lin2016synthetic,di2024quantum,aidelsburger2018artificial} provides opportunities to study both Abelian and non-Abelian structures, as well as associated monopole sources of  different gauge fields. Experiments in synthetic quantum systems have demonstrated $U(1)$ gauge fields with Dirac-monopole solutions \cite{schroer2014measuring,ray2014observation,roushan2014observation}, non-Abelian gauge fields analogous to Yang-Mills theory with Yang-monopole solutions \cite{sugawa2018second,sugawa2021wilson}, and Kalb-Ramond gauge fields with Dixmier-Douady tensor monopole solutions relevant to string theory \cite{tan2021experimental,chen2022synthetic}.

In this work, we extend the study of topological Dirac monopoles to a spin-1 quantum system, where, in contrast to conventional spin-1/2 systems,  topological phases with higher first Chern number and anisotropic synthetic gauge fields are possible~\cite{Hu2018-eo,Palumbo2018-rf}.  Experimentally, the spin-1 configuration is encoded in three microwave-coupled internal electronic states of Bose-condensed rubidium-87 atoms using engineered Hamiltonians that move beyond  spin-vector couplings to include spin-tensor interactions~\cite{stamper2013spinor}. Analogs to such spin-tensor coupled Hamiltonians arise in condensed matter systems, giving rise to spin-tensor Hall currents~\cite{fu2020intrinsic} and bright solitons~\cite{sun2020bright}, exotic Maxwell-fermion quasiparticle excitations in topological semi-metals~\cite{zhu2017emergent,hu2018ttd,lv2017observation,fulga2018geometrically}, and $SU(3)$ topological insulators \cite{barnett20123}. 

Our demonstration of the synthetic anisotropic monopole lends foundational insight into this unique form of matter: we directly measure the monopole's quantized charge and the divergence of its magnetic field near the monopole, and show how varying the parameters of the spin-tensor coupling leads to a topological phase transition between different values of Chern number, characterizing the synthetic charges. Through our ultracold atom platform, we have excellent access to the quantum state that enables measurement of multiple topologically relevant signatures. This work establishes a framework in which to study a broad class of topologically nontrivial synthetic matter systems with  gauge fields beyond the conventional paradigm.

\section*{Gauge theories from geometric and topological structure of spin-1 Hamiltonians}

To begin our discussion of topology and geometric phase in synthetic matter systems, we consider a quantum system governed by a Hamiltonian, which varies according to the values of one or more parameters. At every point in parameter space, the Hamiltonian may be diagonalized and the eigenstates computed. A smooth assignment of eigenspaces over the entire parameter space forms a vector bundle, which has both local (or geometric) and global (or topological) properties. 

Locally, due both to the intrinsic geometric curvature of the system and the arbitrariness of gauge choice, eigenstates may vary point-to-point in parameter space. In the case where the spectrum of the Hamiltonian remains non-degenerate throughout parameter space, the eigenspaces are one-dimensional, so that states may vary by a phase factor, known as the geometric phase. The geometric phase is a powerful tool, as it shares the mathematical structure of electromagnetism: both belong to $U(1)$ gauge theories. Geometric phase is directly analogous to the result of integration of the magnetic vector potential along a path in real space. While the geometric phase is generally gauge dependent, the phase accumulated during adiabatic evolution along a closed loop in parameter space cannot be removed by a gauge transformation; this is referred to as the Berry phase \cite{berry1984quantal,berry1990anticipations,xiao2010berry}. This is directly analogous to the magnetic flux through $\mathcal{S}$, the surface bounded by the loop traversed. Just as the magnetic flux can be written as the closed line integral of the vector potential or the surface integral of the magnetic vector field (or electromagnetic field tensor), Berry phase can be calculated as a surface integral
\begin{equation}
    \gamma =  \int_\mathcal{S}\mathbf{\Berry} \cdot d\mathbf{S}\ = \int_\mathcal{S}\Berry_{\lambda\mu} d{\lambda}d{\mu} \ .
\end{equation}
Here, $\BerryV$ is a pseudo-vector known as the Berry curvature vector, analogous to the magnetic field vector, and $d\mathbf{S}$ is the differential area element normal to the surface $\mathcal{S}$. 
Likewise, $\Berry_{\lambda\mu}$ are components of the Berry curvature tensor in the coordinates on the surface $(\lambda,\mu)$, in analogy to the components of the electromagnetic field tensor. (As in electromagnetism, the tensor and vector forms are related to each other via the Hodge star operation.)

Globally, entire eigenstate bundles can be classified  quantitatively; for complex-valued vector bundles over a two-dimensional parameter space $\mathcal{M}$, the relevant topological invariant is the first Chern number~\cite{chern1946characteristic}
\begin{equation}
    \mathcal{C}_1 = \frac{1}{2\pi}\int_{\mathcal{M}}{\Berry_{\lambda \mu}}d{\lambda}d{\mu} \ ,
\end{equation}
which we simply refer to as the Chern number in this work. This may be understood as the total flux of field lines through an enclosing surface, analogous to magnetic charge. 

In this work, we employ spin-$1$ Hamiltonians, which are elements of $\mathfrak{su}(3)$ and [in contrast to spin-$1/2$ in $\mathfrak{su}(2)$ that is spanned by the three Pauli matrices] require linear combinations of eight basis matrices~\cite{peng2020exotic}; one such choice is the Gell-Mann matrices~\cite{gell2018symmetries}. More convenient here is the basis of spin vectors and spin tensors: the first three spin-vector elements form an $\mathfrak{su}(2)$ sub-algebra, $\hat{\mathbf{F}} = \left(\hat{F}_x, \hat{F}_y, \hat{F}_z\right)$. An additional five matrices to span $\mathfrak{su}(3)$~\cite{wang2020lattice}  are the spin quadrupole tensors~\cite{stamper2013spinor} 
$\hat{N}_{ij} = \frac{1}{2}\left( \hat{F}_i \hat{F}_j +  \hat{F}_j \hat{F}_i  \right)  - \frac{1}{3}\delta_{ij}\hat{\mathbf{F}}\cdot \hat{\mathbf{F}}$.

\begin{figure}
    \centering
    \includegraphics{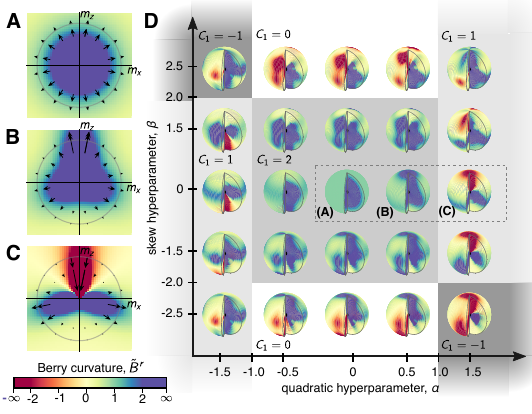}
    \caption{{\bf Berry curvature of the anisotropic monopole.} 
    {\bf A-C.} The Berry curvature $\Berry^r$ is plotted in the $m_x$-$m_z$ plane through the origin for hyperparameter value $\beta = 0$ and $\alpha = 0$ ({A}), $\alpha = 0.5$ ({B}), and $\alpha = 1.5$ ({C}), showing the growth of a ``string'' of positive Berry curvature (blue) from the origin to the north pole as $\alpha$ increases, and the change of sign (red color) as the string inverts direction above the phase transition at $\alpha = 1$. The colorbar indicates $\Berry^r$ values in this and all figures; note that colors are intentionally saturated at $\Berry^r = \pm2$, though curvature values diverge well beyond this. 
    {\bf D.} Mapping Berry curvature over the sphere $r = 1$, with a cut showing internal values, for a variety of hyperparameter values $(\alpha,\beta)$. All cases retain a monopole divergence of $\Berry^r$ at the origin, but the hyperparameter-driven anisotropy results in different structures across the phase diagram. Regions of different topological charge are shaded, with $\mathcal{C}_1$ taking on values of 2, 1, 0, and -1, as indicated.}
    \label{fig:phase_diagram}
\end{figure}

To explore the topological phases in a spin-1 system, first consider the Hamiltonian of simple spin vector couplings $\hat{H} = \hbar\mathbf{m}\cdot\hat{\mathbf{F}}$, where $\mathbf{m} = \left(m_x, m_y, m_z \right)$ is a cartesian parameter space. The synthetic magnetic vector field $\mathbf{\Berry}(\mathbf{m})$ of the ground eigenbundle is isotropic, emanating radially from a monopole at the origin $\left(m_x, m_y, m_z \right) = (0,0,0)$ in parameter space~\cite{Hu2018-eo}. Converting to spherical parameter space coordinates $\mathbf{m}\rightarrow(m_r,m_\theta,m_\phi)$, the synthetic  field is $\BerryV(\mathbf{m}) \propto \mathbf{m}/m_r^3$ (Fig.\ref{fig:phase_diagram}A). Notably, $\BerryV(\mathbf{m}) = (\Berry^r(m_r),0,0)$ always points away from the origin. Integrating the synthetic field over any origin-enclosing two-dimensional surface reveals the Chern number to be  $\mathcal{C}_1=2$. 

The richness of the spin-1 system allows for additional coupling beyond  spin vectors: spin-tensor couplings open up the possibilities for new topological phases and anisotropy of the synthetic fields. Here, we consider two particular instances of spin-tensor coupling~\cite{Hu2018stm, hu2018ttd}:  between $m_z$ and $\hat{F}_z^2$, parameterized by relative strength $\alpha$;  and between $m_x$ and $\hat{N}_{xz}$, parameterized by relative strength $\beta$. The family of Hamiltonians 
\begin{equation}
   {\hat H_{\alpha\beta}}= \hbar\left[\mathbf m \cdot \hat{\mathbf F} + \alpha m_z \hat{F}_z^2  + \beta m_x \hat{N}_{xz}\right] \ 
\label{Eq:Hamiltonian_AB}
\end{equation}
always has, at the origin, a triply degenerate point in the eigenspectrum that is a monopole acting as a source of emanating Berry curvature; closed surfaces enclosing this monopole have well-defined eigenbundles. As the hyper-parameters are tuned to non-zero values, the Berry field becomes anisotropic (Fig.~\ref{fig:phase_diagram}B-D), though it remains that only the radial component $\Berry^r$ is nonzero for all $\alpha$ and $\beta$ (such that $\BerryV(\mathbf{m})$ points towards or away from the monopole at the origin).

In addition to the singularity at the origin, additional degeneracies emerge in the eigenspectrum at critical values of $\alpha$ and $\beta$. Specifically, at  critical hyperparameter values of $\alpha_c= \pm1$ or $\beta_c=\pm2$, degeneracies in the eigenspectrum appear along lines in parameter space emanating from the origin~\cite{hu2018ttd}. These lines of degeneracy pierce any origin-enclosing surface,  causing a band touching in the eigenspectrum and  resulting in an ill-defined Berry curvature. Across these critical hyperparameter values, topological phase transitions between states of different Chern number proceed. Crossing these points, some of the Berry curvature field vectors flip directions, now pointing in towards the monopole (negative $\Berry^r$ in Fig.~\ref{fig:phase_diagram}C), and the Chern number, being the total flux of field vectors through an enclosing surface, changes value. By changing  $(\alpha,\beta)$, topological phases with Chern numbers $\mathcal{C}_1 = 2, 1,0,-1$ can all be reached. As hyperparameters  change, not only can we  simulate magnetic monopoles of different synthetic charge, we also deform the synthetic field emanating from the monopole towards anisotropy in parameter space (Fig.~\ref{fig:phase_diagram}D).

\section*{Engineering the Hamiltonian}

To synthesize a tensor-coupled spin-1 Hamiltonian of Eq.~\ref{Eq:Hamiltonian_AB}, we use $\approx10^5$ Bose-Einstein condensed rubidium-87 atoms acting as an ensemble of identical, non-interacting quantum systems~\cite{supmat}. Three states $\{\ket{1},\ket{2},\ket{3} \}$ from the  hyperfine ground-state manifold form the pseudo-spin-1 system, which are coupled via two independent microwave fields~\cite{supmat}. The three-level system is governed by the effective Hamiltonian
 \begin{equation}
  \label{drive Ham 2}
    \hat{H} = \hbar
    \begin{pmatrix}
        \Delta_1 & \frac{\Omega_{1}}{2}e^{i\phi_1} & 0 \\
        \frac{\Omega_{1}}{2}e^{-i\phi_1} & 0 & \frac{\Omega_{2}}{2}e^{i\phi_2} \\
        0 & \frac{\Omega_{2}}{2}e^{-i\phi_2}&  \Delta_2
    \end{pmatrix},
\end{equation}
where Rabi frequencies $\Omega_{1,2}$, phases $\phi_{1,2}$, and  detunings from resonance $\Delta_{1,2}$ are carefully controlled parameters of the microwave fields. Though complete control over this six-parameter hamiltonian is possible~\cite{lindon2023cuq}, we reduce our system to the three-parameter space~\cite{Hu2018stm, Palumbo2018-rf} [characterized in the parameter space spanned by $\mathbf{m} = (m_r,m_\theta, m_\phi)$] by constraining the experimental controls as
\begin{align}
         \Omega_1 &= 2{m_r\sin{m_\theta}}\sqrt{\left[ (1+\beta/2)^2 \cos^2 m_\phi + \sin^2 m_\phi\right]/2}\\
        \Omega_2 &= 2{m_r\sin{m_\theta}}\sqrt{\left[ (1-\beta/2)^2 \cos^2 m_\phi + \sin^2 m_\phi\right]/2}\\  
        \Delta_1 &= m_r(\alpha +1)\cos{m_\theta} , \ \Delta_2 = m_r(\alpha -1)\cos{m_\theta}\\
        \phi_1 &= -\arctan \left(\frac{\tan m_\phi}{1+\beta/2}\right) , \ 
        \phi_2 = -\arctan \left(\frac{\tan m_\phi}{1-\beta/2}\right).
    \label{Eq:ParameterMap}
\end{align}
Under this mapping, Eq.~\ref{drive Ham 2} reduces to Eq.~\eqref{Eq:Hamiltonian_AB}, and hyperparameters $\alpha$ and $\beta$ control the contributions of the spin tensor terms in the simulation.

\section*{Experimentally measuring Berry curvature}

Quantitative characterization of the topological phases requires Berry curvature measurements at points across the parameter space, which, when integrated, reveal $\mathcal{C}_1$ and the topological charge. We use slightly non-adiabatic evolution~\cite{gritsev2012dqh,schroer2014measuring,sugawa2018second,roushan2014observation}, where the first-order correction to adiabatic evolution is related to the Berry curvature at any point in the parameter space, as long as no gaps close. The principle underlying this method is that the ``error'' in a measured state found after non-adiabatic evolution is proportional to the variation in eigenstates along the trajectory taken, which is exactly the measure the Berry curvature vector accounts for.  For a small degree of nonadiabiticity, a generalized force in the $\mu$-direction $\langle \mathcal{F}_{\mu} \rangle = -\langle \psi | \partial_{\mu}\hat{H} | \psi \rangle \approx \mathrm{const} - v_{\lambda}\Berry_{\lambda \mu} \ $  arises due to a change in parameter $\lambda$ with speed $v_{\lambda} = {d\lambda}/{dt}$. For the spin-1 Hamiltonians of Eq.~\ref{Eq:Hamiltonian_AB}, 
\begin{equation}
    \begin{split}
         \langle  \mathcal{F}_\phi \rangle = m_r\big(\sin m_\theta \sin m_\phi \langle \hat F_x\rangle &-  \sin m_\theta \cos m_\phi \langle \hat F_y\rangle \\ &+\beta \sin m_\theta \sin m_\phi \langle \hat N_{xz}\rangle\big)
    \end{split}
\end{equation}

In practice, $\Berry_{\theta\phi}$ may be found by ramping $\theta$ at different values of $\phi$ and recording the resulting spin expectation values $\langle \hat{F}_{x} \rangle$, $\langle \hat{F}_{y} \rangle$, $ \langle \hat N_{xz}\rangle$, so long as $v_\theta$ gives a rate in the nearly adiabatic regime: $\Omega_{1,2}^2 \gg v_\theta^2$. For studies with all $\beta = 0$, the constant zeroth-order contribution vanishes, and to leading order
\begin{equation}
    \Berry_{\theta\phi} = -\frac{m_r}{v_\theta}\left[ \sin m_\theta \sin m_\phi \langle \hat{F}_x \rangle - \sin m_\theta \cos m_\phi \langle \hat{F}_y \rangle \right].
    \label{eq:ramp}
\end{equation}
The Berry curvature vector, being the Hodge dual of the curvature tensor, in spherical coordinates has a radial component 
\begin{equation}
    \Berry^r = \frac{\Berry_{\theta \phi}}{m_r^2 \sin m_\theta}  .
\end{equation}
Additionally, the $\beta = 0$ Hamiltonians have azimuthal symmetry and thus the Berry curvature $\Berry_{\theta\phi}$ is independent of $m_\phi$. Measuring  along a single line of longitude suffices to determine the curvature over the entire parameter space in this case, and 
\begin{equation}
    \label{Eq: experimental chern}
    C_1 = \int_0^\pi \Berry_{\theta 0} \, d m_\theta = \int_0^T m_r \sin m_\theta(t) \expect{\hat{F}_y(t)}~dt.
\end{equation}

\section*{Monopole singularity and topological protection}

\begin{figure}
    \centering
    \includegraphics{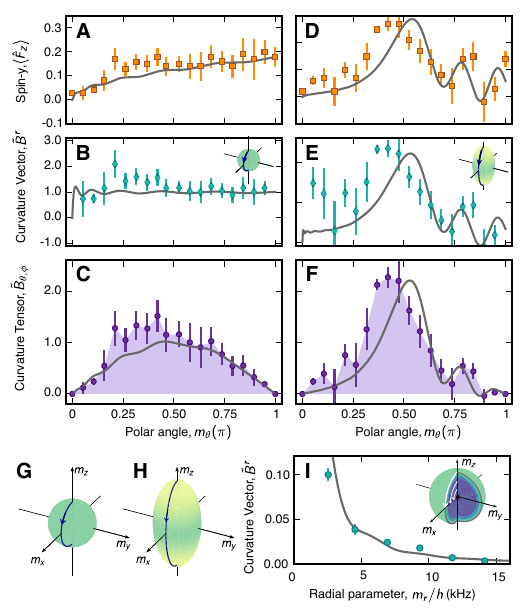}
    \caption{\textbf{Characterizing the synthetic monopole}. 
    In the case of the  uniform monopole $(\alpha, \beta) = (0,0)$, the method of not-quite-adiabatic evolution is used both along the path $r =1$ {\bf(A-C)} and from a deformed path from $m_z = 2$ to $-2$ via $m_x = 1$ {\bf(D-F)}.  
    Spin projections $\expect{\hat{F}_y}$ {(A, D)} are measured vs.\ final polar angle $m_\theta$, from which the curvature vector $\Berry^r$ {(B, E)} and tensor $\Berry_{\theta\phi}$ {(C, F)} are determined. Calculations of the evolution using all experimental parameters are shown via grey lines in all subfigures; uncertainties shown reflect variation over repeated measurements. The insets in {(B, E)} illustrate the path used for measurement and the calculated $\Berry^r$ on the parameter-space sphere. 
    {\bf (G)} Spherical path through parameter space for (A-C) measurements, superimposed on sphere representing $\Berry_r(\mathbf{m})$ (colorscale as in Fig.~\ref{fig:phase_diagram}). {\bf (H)} Deformed path through parameter space for (D-F) measurements together with $\Berry_r(\mathbf{m})$; 
    {\bf (I)} Using the same measurement scheme as in {(A-C)}, $\Berry^r$ is measured at different distances from the monopole, indicating the divergence of this field at the origin. Solid line is a calculation using experimental parameters. Inset shows three representative paths in parameter space. }
    \label{Fig:MonopoleCharacter}
\end{figure}

Two characteristic features of an electromagnetic monopole are that the flux through any closed surface surrounding it remains constant regardless of the surface’s shape, and that the field strength decreases with the inverse square of the distance from the monopole. We experimentally demonstrate the topological protection of the Chern number (i.e. synthetic monopole charge) under  hyperparameters ($\alpha=0$, $\beta=0$), which  corresponds to the isotropic $\BerryV(\mathbf{m}) = (\Berry^r(m_r), 0, 0)$ field. First, we measure $\Berry^r$ over a spherical shell of $m_r = 1$. 
Next, we deform the effective shell of integration by stretching the $m_z$-component of parameter space relative to the $m_x$ and $m_y$ components in a way that effectively keeps $\BerryV(\mathbf{m})$ the same in parameter space, but makes the  $(m_\theta,m_\phi)$ parameters sweep out an ellipsoidal surface of integration. While   $\BerryV(\mathbf{m})$ is  constant everywhere on spherical shells, it is not constant over an ellipsoidal shell (Fig.~\ref{Fig:MonopoleCharacter}(G)).

We begin by preparing the ground eigenstate under north pole parameters $(m_r, m_\theta = 0,m_\phi)$, which corresponds to simple uncoupled Hamiltonians with Rabi frequencies $\Omega_1 = \Omega_2 =0$. We vary the polar angle $m_\theta$ along the line of longitude $m_\phi = 0$ with time-dependence $m_\theta(t) = a t^2 $ where $a = {\pi}/{T^2}$ and $m_\theta(T) = \pi$ at the end of the ramp~\cite{Hu2018stm}. In repeated iterations of the experiment, we stop the evolution at twenty  evenly spaced angles along this path and measure  $\expect{\hat{F}_y}$ via quantum state tomography (QST)~\cite{supmat}. We use Eq.~\ref{eq:ramp} to calculate Berry curvature $\Berry_{\theta 0}$ and numerically compute the integral Eq.~\eqref{Eq: experimental chern} to find the topological charge $\mathcal{C}_1$.

Figure~\ref{Fig:MonopoleCharacter} shows measured spin components$\langle\hat{F}_y\rangle$, Berry curvature  $\Berry^r$, and tensor $\Berry_{\theta\phi}$  throughout the ramps for both the spherical and elliptically deformed paths. While $\Berry_{\theta\phi}$ clearly differs as a function of $\theta$ between the spherical and ellipsoidal trajectory, the integrals characterizing the sum of all flux lines through the surfaces are both found to be $C_1 = 2.5 \pm 0.2$. This is found from the data in Figs.~\ref{Fig:MonopoleCharacter}C and F, where the curve shapes differ but the area under the curve traced out by $\Berry_{\theta \phi}$ is the same for both paths. 
While the results show a small quantitative deviation from the predicted $C_1 = 2$ value here, the mutual agreement between the two measurements is strong qualitative evidence for the expected independence from the integration volume. The residual is consistent with a systematic offset arising from the combination of stringent calibration requirements and propagation of small offsets through two layers of calculation.
Overall, these measurements reveal a clear signature of topological invariance in this synthetic monopole: the Chern number, which classifies the topological charge of the Hamiltonian's ground state eigenbundle, is invariant to deformations. 

A second signature of the monopole is revealed by measuring $\BerryV(\mathbf{m})$ over spherical shells of varying radius. The Berry curvature $\BerryV(\mathbf{m})$ is analogous to the magnetic field vector that diverges near a monopole source, and we observe the expected inverse-square scaling by fixing the parameters ($m_\theta, m_\phi$)  and varying  the radius $m_r$. We choose a point on the equator ($\theta = \pi/2$, $m_\phi = 0$) and measure $\expect{\hat{F}_y}$ using not-quite-adiabatic ramps. Figure~\ref{Fig:MonopoleCharacter}G shows the  $\Berry^r$ as functions of radius, and its divergence as $m_r\rightarrow 0$, an experimental signature of the topological monopole and the synthetic  charge at the origin of parameter space.

\section*{Directly measuring a topological phase transition}

Having characterized the key features of the synthetic monopole, we delve into studies of its anisotropic nature under the introduction of spin-tensor coupling.
As the hyperparameters  vary, not only does the $\BerryV(\mathbf{m})$ become anisotropic, but we also find topological phase transitions associated with changes in the monopole charge. 

We experimentally study Hamiltonians with $\alpha$ ranging from 0 to 2, crossing a critical value $\alpha = 1$. Due to the azimuthal symmetry with $\beta=0$,  we extract the Chern number by integrating  over a single line of longitude (Eq.~\ref{Eq: experimental chern}). We again use the technique of nearly adiabatic evolution, measuring $\langle \hat{F}_y\rangle$ as a function of polar angle $\theta$ to extract the Berry curvature. 

\begin{figure*}
    \centering
    \includegraphics{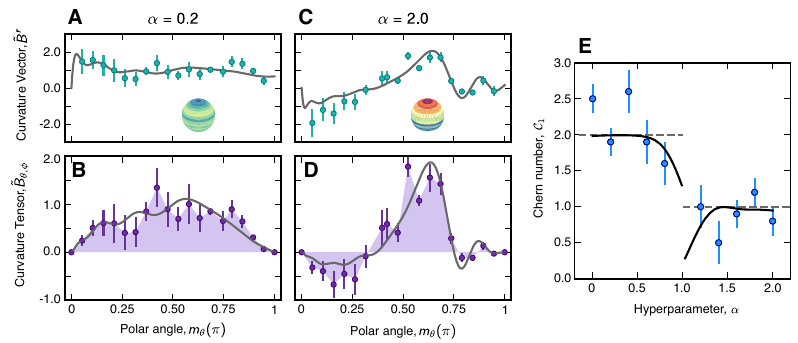}
    \caption{\textbf{Topological phase transition of anisotropic monopole}. By measuring the Berry curvature vector $\Berry^r$ {\bf(A,C)} and tensor $\Berry_{\theta\phi}$ {\bf(B,D)} values over the polar angle $m_\theta$ for $(\alpha,\beta$) = (0.2,0) {(A-B)} and  (2.0,0) {(C-D)}, the first Chern number is obtained via numerical integration of $\Berry_{\theta\phi}$. Insets to {\bf(A,C)} show  $\Berry^r$ over the $m_r = 1$ parameter space sphere using the colorscale from Fig.~\ref{fig:phase_diagram}. Uncertainties shown reflect variation over repeated measurements. {\bf (E)} Chern number $\mathcal{C}_1$ measured vs.\ $\alpha$ (round points), indicating the phase transition at $\alpha = 1$. Solid line shows the calculation including experimental parameters, while dashed line indicates the idealized expectation  of infinite-time measurement.} \label{Fig:ExperimentalBerry}
\end{figure*}

As we vary the spin tensor component  with $\alpha = 0$ to $\alpha = 2$, we see distinct changes in  the Berry curvature and the resulting Chern numbers  on opposite sides of the $C_1=2 \leftrightarrow 1$ transition at $\alpha =1$. For $\alpha = 0.2$ and $\alpha = 2.0$, we measure the Berry curvature vector values $\Berry^{r}(1,m_\theta,0)$ along the $m_\theta$-path. While both have some degree of anisotropy (in contrast to the $\alpha = 0$ case of Fig.~\ref{Fig:MonopoleCharacter}B), the magnitude of all  field values are positive for $\alpha = 0.2$ (Fig.~\ref{Fig:ExperimentalBerry}A), but some are negative for $\alpha = 2.0$, indicating that $\BerryV(\mathbf{m})$ points inwards (Fig.~\ref{Fig:ExperimentalBerry}C). Fig.~\ref{Fig:ExperimentalBerry}B and Fig.~\ref{Fig:ExperimentalBerry}D show the corresponding Berry curvature tensor element $\Berry_{\theta, 0}$. Numerical integration of these data yields two distinct Chern numbers: $C_1 = 1.9 \pm 0.2$ and $C_1 = 0.8 \pm 0.2$ for $\alpha = 0.2$ and $2$, respectively.

By repeating this measurement procedure for ten different spin-tensor contributions ($\alpha$-values), we observe a distinct change in the Chern number across the phase transition at $\alpha = 1$ (Fig.~\ref{Fig:ExperimentalBerry}E), revealing the  topological phase transition in the synthetic spin-1 monopole systems.

\section*{Signatures of topological phases}
\begin{figure}
    \centering
    \includegraphics{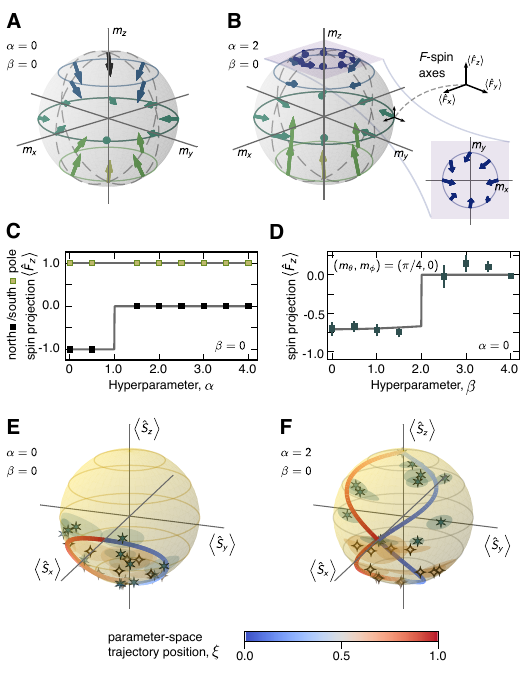}
    \caption{{\bf Complementary signatures of the anisotropic monopole.} {\bf (A,B)} Spin-projection measurements of $\mathbf{F}$ for systems adiabatically prepared across {\bf m}-parameter space for $\alpha = 0$ ({A}) and $\alpha = 2$ ({B}) and $\beta = 0$. Each arrow represents the orientation of {$\mathbf{F}$} in a set of spin axes centred at the arrow tail (orientation as shown, right). Outset from {(B)} is the projection in the $m_x$-$m_y$-plane at $m_\theta = 0.436$~rad $(=25^\circ)$. {\bf (C)} The $\expect{F_z}$ projection measurement  at the north (${\bf m} = (0,0,m_z = 1)$, black) and south (${\bf m} = (0,0,m_z =-1)$, green) poles as a function of $\alpha$ for $\beta = 0$. Error bars are smaller than data points. {\bf (D)} The $\expect{F_z}$ projection measurement at the north [${\bf m} = (0,0,m_z =1/\sqrt{2})$ or $(m_\theta,m_\phi) = (\pi/4,0)$] pole as a function of $\beta$ for $\alpha = 0$. Error bars represent statistical variations.  {\bf (E, F)} Majorana star projections on the Bloch sphere for systems adiabatically prepared along the path [Eq.~\eqref{Eq:trajectory}] through {\bf m}-parameter space. for $\alpha = 0$ ({E}) and $\alpha = 2$ ({F}) and $\beta = 0$. The two stars are represented as blue six-point and orange four-point stars; statistical experimental uncertainties are represented by shaded ellipses. Theoretical star locations represented as the colored curves, with the trajectory position (parameterized by $\xi$) indicated as per the colorbar. In both cases, two trajectories are present: in (E), they lie on top of each other, but in (F) lie in opposite hemispheres.}
    \label{fig:Signatures}
\end{figure}

While directly computing the Chern number $C_1$ gives quantitatively distinguishing information for  different topological phases, signatures of these transitions are often found in other metrics. For spin-1 systems, the three-component quantum state  lacks the simple geometric mapping to the Bloch sphere found for spin-$1/2$, meaning that the geometric phase and the Chern number cannot be inferred directly from intuitive visual features like the winding of a state enclosing a solid angle~\cite{hannay1998berry}, nor can tracking the spin vector $\langle \hat{\mathbf{F}} \rangle$ reveal the full information of the quantum state as it does in the two-state case. However,  some qualitative visual cues do herald a topological phase transition in a three-level system~\cite{Hu2018stm}, including in the overall spin vector orientation across the  parameter space, sometimes known collectively as the spin texture~\cite{petti2022review}: spin-texture vortices appear and disappear as topological phase boundaries are crossed, as demonstrated in Fig.~\ref{fig:Signatures}.

To measure spin-texture transitions between systems with hyperparameters $\alpha$ and $\beta$ defining different topological phases, we adiabatically prepare systems in the ground states with particular ($\alpha,\beta$) and perform quantum state tomography (QST) to measure $\langle\hat{\mathbf{F}}\rangle$ across $m$-parameter space. 

We prepare systems at the north pole $(m_r, m_\theta = 0,m_\phi)$. Here, for $\alpha > 1$ and $\alpha < 1$, the ground states are bare atomic eigenstates: for $\alpha<1$, the ground state is $| \psi \rangle = | 3 \rangle$; for $\alpha>1$ it is $| \psi \rangle = | 2 \rangle$ (the eigenstates here do not depend on $\beta$). We then perform adiabatic rapid passage by linearly ramping $\Omega_{1,2}, \Delta_{1,2}, \phi_{1,2}$ in $1.0$~ms to the target point in parameter space (Fig.~\ref{fig:Signatures}A,B). Our calculations~\cite{supmat} guarantee this ramp duration is sufficiently long to suppress non-adiabatic effects and allow high-fidelity state preparation, while still short enough that ambient-field driven detunings did not cause significant noise in the final reconstructed spin vector.

For ($\alpha =0,\beta = 0$), the measured spin vectors all point radially towards the monopole at $r = 0$. (Fig.~\ref{fig:Signatures}A). For ($\alpha =2,\beta = 0$), we observe the emergence of a vortex in the spin texture surrounding the north pole, corresponding the expected thread of degeneracies in the eigenspectrum from
the origin along positive-$m_z$. The spin vector directly at the pole vanishes, and the spin vectors surrounding the pole circulate around the centre of the vortex (Fig.~\ref{fig:Signatures}B). To emphasize this effect, we measure spin vectors on an additional ring displaced $m_\theta = 25^{\circ}$ from the pole. Plotting this slice from a top-down view (Fig.~\ref{fig:Signatures}B, outset), the spin vectors all approximately point radially towards the vortex core. 

This phase transition is also revealed in the observable $\langle\hat{F}_z\rangle$ at the north pole: there is a discrete jump at $\alpha = 1.0$ (Fig.~\ref{fig:Signatures}C). Likewise, measuring $\langle\hat{F}_z\rangle$ at $(m_\theta = \pi/4, m_\phi = 0)$ as a function of $\beta$ reveals the  $C_1 = 2 \leftrightarrow 0$ phase transition (Fig.~\ref{fig:Signatures}D). 

An additional signature for distinguishing topological phases can be found by visualizing states using the Majorana stellar representation (MSR)~\cite{serrano2020majorana}. The MSR extends the idea of stereographically projecting a qubit state onto the Bloch sphere: while a general spin-$s$ state can not be represented as a point on a sphere, 
the MSR represents the state by $2s$ points on a sphere. For a general spin-$s$ state expanded in the $\hat{S}_z$ eigenbasis, $|\psi\rangle = \sum_{m=-s}^{s}\lambda_m |sm \rangle$, the $2s$ complex numbers to be stereographically projected onto the sphere are the roots $\{\zeta_k = \tan(\theta_k/2) e^{i\phi_k}\}_k$ of the Majorana polynomial
\begin{equation}
    p_\psi(Z) = \sum_{m=-s}^{s} (-1)^{s-m} \sqrt{ \binom{2s}{s-m}} \lambda_m Z^{s+m} \ .
    \label{Eq:MajPoly}
\end{equation}
The trajectories of these stars depend sensitively on the topological phase. For the same path through the base parameter space, the stars trace out significantly different motions. We adiabatically prepare the ground state and perform QST to experimentally determine the locations of the Majorana stars at multiple points along a fixed-$r$ path through parameter space~\cite{Hu2018stm},
\begin{align}
        \theta(\xi) &= \frac{\pi}{4}\cos\left({2\pi \xi}\right) +\frac{\pi}{4}, \\ 
        \phi(\xi) &= \frac{\pi}{3}\sin\left({2\pi \xi}\right),
    \label{Eq:trajectory}
\end{align}

where $\xi$ takes values from $(0,1)$.
 As shown in Fig.~\ref{fig:Signatures}E, for ($\alpha =0,\beta = 0$), the two stars move approximately together in a loop bounded to the southern hemisphere of the Bloch sphere. However, as shown in Fig.~\ref{fig:Signatures}F, for ($\alpha =2,\beta = 0$), the stars trajectories dramatically separate. One star moves from the equator to the north pole, while the other from the equator to the south pole. 

\section*{Discussion and conclusions}

This work demonstrates a quantum simulator of a synthetic spin-1 monopole that can take on different non-zero charges, which are characterized by the first Chern number. The spin-tensor coupling that controls this monopole charge simultaneously tunes the anisotropic geometric structure of the associated {synthetic magnetic} field, characterized by the Berry curvature, demonstrating the intrinsic relationship between geometry and topology in this system~\cite{Mera2022-nf}. The critical points where {the curvature becomes singular} coincide with  topological phase transitions between systems of different, quantized monopole charge.  Our measurements of Berry curvature, spin textures, and Majorana star trajectories all provide signatures of the parameter-dependent topology and geometry in this ultracold atom experiment.

Synthetic quantum matter with spin-tensor–momentum interactions provides a rich playground for exploring unique topological phases in high-dimensional systems. This work adds monopole sources of anisotropic mutagenic {magnetic} fields to the toolbox of artificial gauge theory simulations, with a promising outlook towards richer phases in higher-spin Hamiltonians, enabled through additional independently controlled microwave couplings in atomic systems like the one presented here. Extending this work by implementing real-space landscapes through which the Hamiltonian changes~\cite{dalibard2011colloquium, goldman2014light,Grass2025-yc} could enable studies of quantum dynamics in high-spin systems subject to spin-tensor couplings. Additionally, recent developments in single-atom control~\cite{Labuhn2014-mw,Shaw2024-pe,Wang2025-iz} offer opportunities for individually addressing atoms' topological phases in many-particle systems, and introducing many-body interactions between particles may enable increasingly complex phases of synthetic matter.

\begin{acknowledgments}
The authors acknowledge preliminary experimental contributions from Logan W. Cooke, Joseph Lindon, and Anna Prus-Czarnecka, especially in apparatus stability improvements that made this work possible. As researchers at the University of Alberta, we acknowledge that this work was conducted on Treaty 6
territory, and that we respect the histories, languages, and cultures of First Nations, M\'etis, Inuit, and all First Peoples of Canada, whose presence continues to enrich our vibrant community.
\end{acknowledgments}

\paragraph*{Funding:}
This work is supported by funding sources NSERC (RGPIN-2021-02884 and ALLRP 578462-22, ALLRP 578468-22, ALLRP 578460-22), 
Canada Foundation for Innovation (36130), the Canada Research Chairs (CRC) Program, Alberta Innovates, the Province
of Alberta, and the University of Alberta. A.T. is funded by the Austrian Science Fund (FWF) 10.55776/RIC1995124.

\paragraph*{Author contributions:} 
N.M.\ and A.T.\ contributed equally to this work. 
N.M.\ performed experiments, theoretical modelling,  formal analysis, writing (original draft); 
A.T.\ performed conceptualization, theoretical modelling, formal analysis, visualization, and writing (review and editing); 
K.T.\ assisted with experiments and visualization;
D.F. assisted with experiments; and
L.J.L.\ supervised the work, acquired funding, and  performed visualization and writing (review and editing).

\appendix
\section{Defining the spin-1, three-level system}
We choose three states from rubidium-87's hyperfine ground-state $5S_{1/2}$ manifold:   
$|{f=2,m_f=2}\rangle \equiv |1\rangle$, $ |{f=1,m_f=1} \rangle \equiv |2\rangle$, and $|{f=2,m_f=1} \rangle\equiv |3\rangle$
to form the pseudo-spin-1 system (Fig.~\ref{fig:S1}). 
An external magnetic field splits states within the same total angular momentum $f$ via the Zeeman effect, resulting in energetic differences of $\hbar\delta_z=2\pi \times 3.9$~MHz between adjacent sublevels. The three states are coupled via two independent microwave fields (Fig.~\ref{fig:S1}(A)), each characterized by three parameters: the field strength, characterized by Rabi frequencies $\Omega_{1,2}$;  phases $\phi_{1,2}$; and  central frequencies (i.e. detunings from resonance) $\Delta_{1,2}$. In the frame rotating at the drive frequencies, this coupling scheme yields the effective Hamiltonian given in Eq.~\ref{Eq:Hamiltonian_AB}.

\section{Experimental sequence}

\begin{figure}
    \centering
    \includegraphics{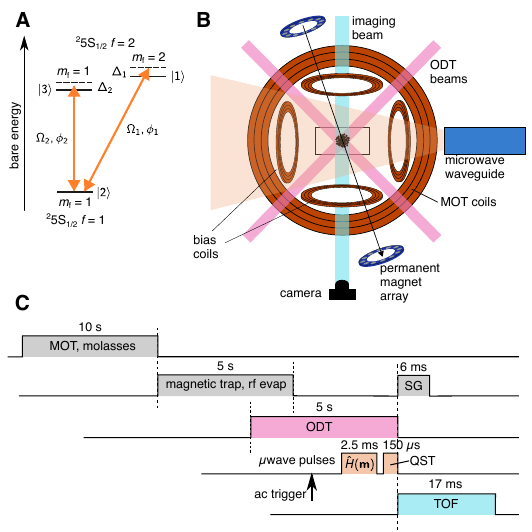}
    \caption{\textbf{Experimental diagram}. (\textbf{A}) Level diagram of three levels we select for spin-1 system. (\textbf{B}) Experimental diagram simplified: the ultracold atomic cloud is prepared in an ODT. Permanent magnets create a Zeeman splitting. An open-ended waveguide irradiates the atoms with microwaves. (\textbf{C}) Time-ordering of experimental sequence (not to scale): MOT is magneto-optical trap; MT is magnetic trap; rf evap is radiofrequency evaporation, ODT is optical dipole trap; QST is quantum state tomography; SG is Stern Gerlach separation; TOF is time of flight. }
    \label{fig:S1}
\end{figure}

The experimental apparatus for generating Bose-Einstein condensate (BECs) of rubidium-87 atoms (Fig.~\ref{fig:S1}(B)), and typical methods, are described in detail in our previous work~\cite{ saglamyurek2021storing, lindon2023cuq, cooke2024ife,milson2023high}. Our sequence, shown  in Fig.~\ref{fig:S1}(C), is repeated many times to collect statistics and vary parameters for each of our measurements. In short, each sequence begins with laser cooling and trapping $^{87}$Rb atoms using a both a 2-dimensional and 3-dimensional magneto-optical trap (MOT) for 10~s,  polarization gradient sub-doppler cooling for 20~ms, forced RF evaporation from a magnetic trap for 5~s, and a crossed beam optical dipole trap (ODT) for 5~s. Each sequence, we  prepare an ensemble of approximately $10^5$ $^{87}$Rb ultracold atoms at approximately 350~nK.

Next, we trigger on the AC line and wait for the same phase of the 60~Hz signal before proceeding, to ensure the atoms are subject to the same electromagnetic background field for the next phase of each cycle.  Microwave radiation is then delivered to the ensemble for a desired duration using an open-ended waveguide resulting in an effectively unpolarized beam. The input signal is generated by amplitude modulating a microwave carrier with a dual-tone baseband signal from an arbitrary waveform generator (AWG). A sample magnetic field reaching the atoms, of the relevant sideband near resonance, with dual tone frequencies $\omega_{1}$ and $\omega_2$, is $\mathbf{B}(t) = b_x\left[A\cos(\omega_1t + \phi_1) + B\cos(\omega_2t + \phi_2) \right]\mathbf{e}_x + b_z\left[ A\cos(\omega_1t + \phi_1) + B\cos(\omega_2t + \phi_2) \right]\mathbf{e}_z$, where $A$ and $B$ are the tone amplitudes (which are proportional to the Rabi frequencies $\Omega_1$ and $\Omega_2$, and $b_x$ and $b_z$ describe the polarization fraction in each of the two directions in the plane transverse to propagation. The detunings are therefore $\Delta_{1} = \omega_1 - \omega_{12}$ and $\Delta_{2} = \omega_2 - \omega_{32}$, where $\omega_{12}$ is the resonance frequency for the transition between $\ket{1}$ and $\ket{2}$, and $\omega_{32}$ for $\ket{3}$ and $\ket{2}$.

Before the main microwave sequence, the atoms all begin in the $|1 \rangle$ state, as this is the state suspended by the magnetic trap. If, instead, states $|2\rangle$ or $|3\rangle$ are desired as the initial states, the atoms are adiabatically transferred by applying a fixed microwave tone while sweeping a bias magnetic field.

\begin{figure}
    \centering
    \includegraphics[width=1\linewidth]{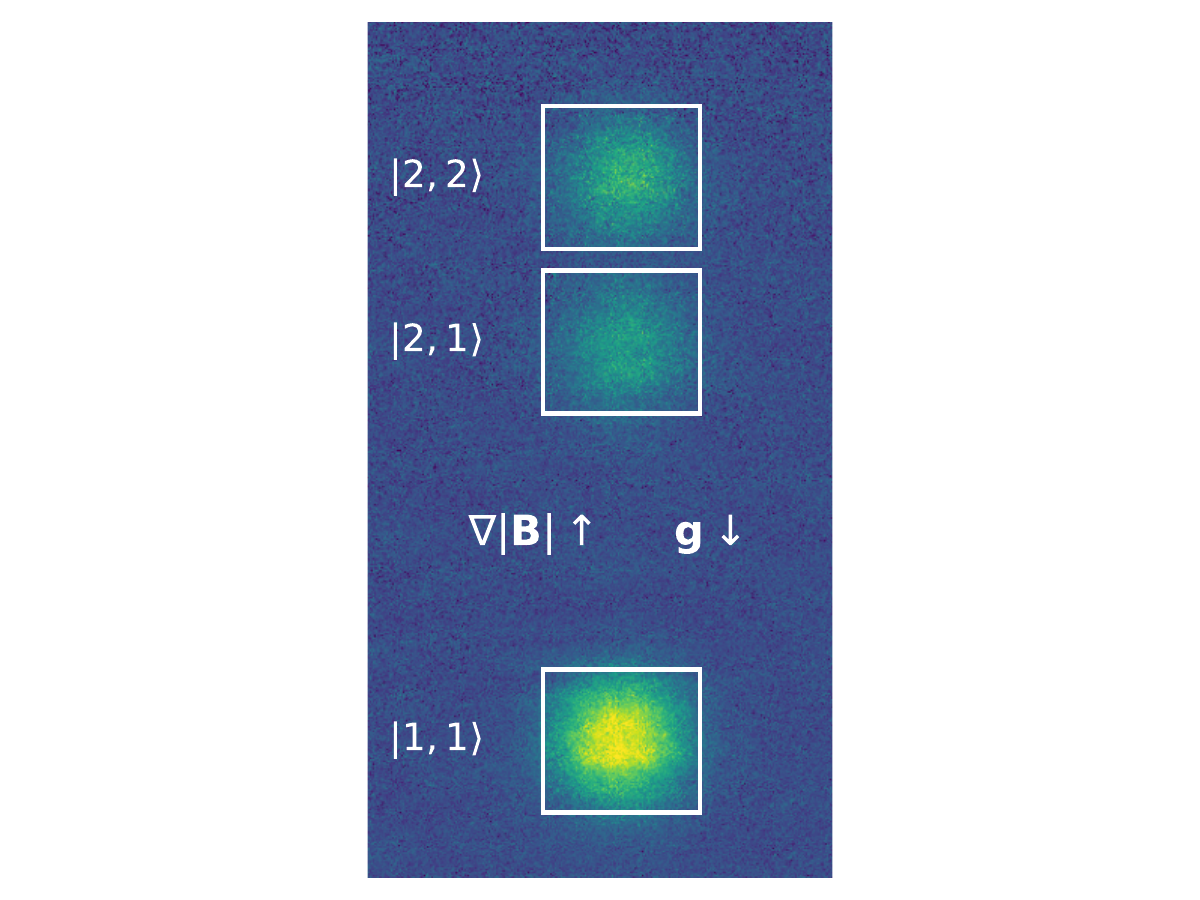}
        \caption{\textbf{TOF absorption image}. A sample TOF absorption image. The three relevant states are split via a Stern-Gerlach technique. The directions of the magnetic field gradient and gravity are indicated.}
        \label{fig:S2}
\end{figure}

After executing a sequence of microwave pulses, we determine the populations in each state of the bare atomic basis $\{\ket{1}, \ket{2}, \ket{3} \}$ using a Stern-Gerlach-like measurement. Within the two ground hyperfine manifolds, the chosen magnetic sublevels all have distinct $z$-axis magnetic moments proportional to the product of their magnetic quantum number and Land\'e g-factor. As we release atoms from the ODT, a magnetic field gradient is applied during time-of-flight (TOF) and exerts a force on the atoms causing spatial separation of the atomic clouds according to their magnetic moments. To create the necessary field gradient, we repurpose the same anti-Helmholtz coil pair used for the MOT and magnetic trap. The atoms fall through a gradient of $27~ \mathrm{G}/\mathrm{cm}$ for 6~ms, followed by an additional 11~ms of TOF. After 17~ms, the ensemble has spatially separated into three distinct clouds corresponding to the different magnetic moments.  We illuminate the atoms using probe light resonant with the D2 $f=2 \leftrightarrow f^\prime =3$ transition and  capture an absorption image. To ensure that the atoms in the $f=1$ state are  imageable,  we pump the $f=1$ atoms into the $f=2$ manifold with a D2 $f=1 \leftrightarrow f^\prime =2$ resonant beam after TOF  separation but before illumination with the imaging probe. The resulting shadow image, recorded using a CCD camera, allows us to estimate the atom number in each sub-cloud and thus infer the population distribution across the three states. A sample absorption image is shown in Fig.~\ref{fig:S2}.

\section{Quantum state tomography}

Our absorption imaging method implements a projective measurement of all three states in the bare atomic Hamiltonian's ($\hat{F}_z$) eigenbasis, which we refer to as our computational basis. Because the atomic ensemble is dilute and approximately non-interacting, the Stern-Gerlach TOF measurement acts as a simultaneous projective measurement on all $\sim 10^5$ qutrits. Each shot yields the relative population of the three computational basis states, corresponding to the square moduli of the generally complex amplitudes. However, the populations determined by the Stern-Gerlach absorption image technique are insufficient to reconstruct the full quantum state, which also requires relative phase information. Full spin-1 quantum state tomography (QST) is therefore necessary.

Our tomographic sequence is described in detail in \cite{lindon2023cuq}. For each experimental pulse sequence effecting a particular parameter set, we repeat the QST sequence six times. Each of these QST sequences involves different additional ``readout" unitary operator $\hat{R}_i$ before the projective measurement. These $\hat{R}_i$ are chosen to effectively rotate the state into different measurement bases. That is to say, applying a sequence of pulses followed by a projective measurement in the computational basis like
\begin{equation}
    \left|\langle j |\hat{R}_i\hat{U} |\psi_0 \rangle\right|^2
\end{equation}
where $\hat{U}$ is the pulse sequence of interest applied to initial state $| \psi_0 \rangle$, and $| j \rangle, \ j=1,2,3$ are the computational basis states. Therefore, this can equivalently be thought of as a measurement of $\hat{U} | \psi_0 \rangle$ in the $\hat{R_i}^{\dagger} | j \rangle$ basis. 

The specific readout unitaries are listed in Ref.~\cite{lindon2023cuq}, along with details on how expectation values of the orthogonal Gell-Mann matrices needed to fully reconstruct the qutrit density matrix can be derived from the resulting population measurements. The final state is then reconstructed using an iterative maximum likelihood estimation technique~\cite{hradil1997quantum,lvovsky2004iterative}.

\section{Calibrating Rabi frequency and detuning}

\begin{figure}
    \centering
    \includegraphics[]{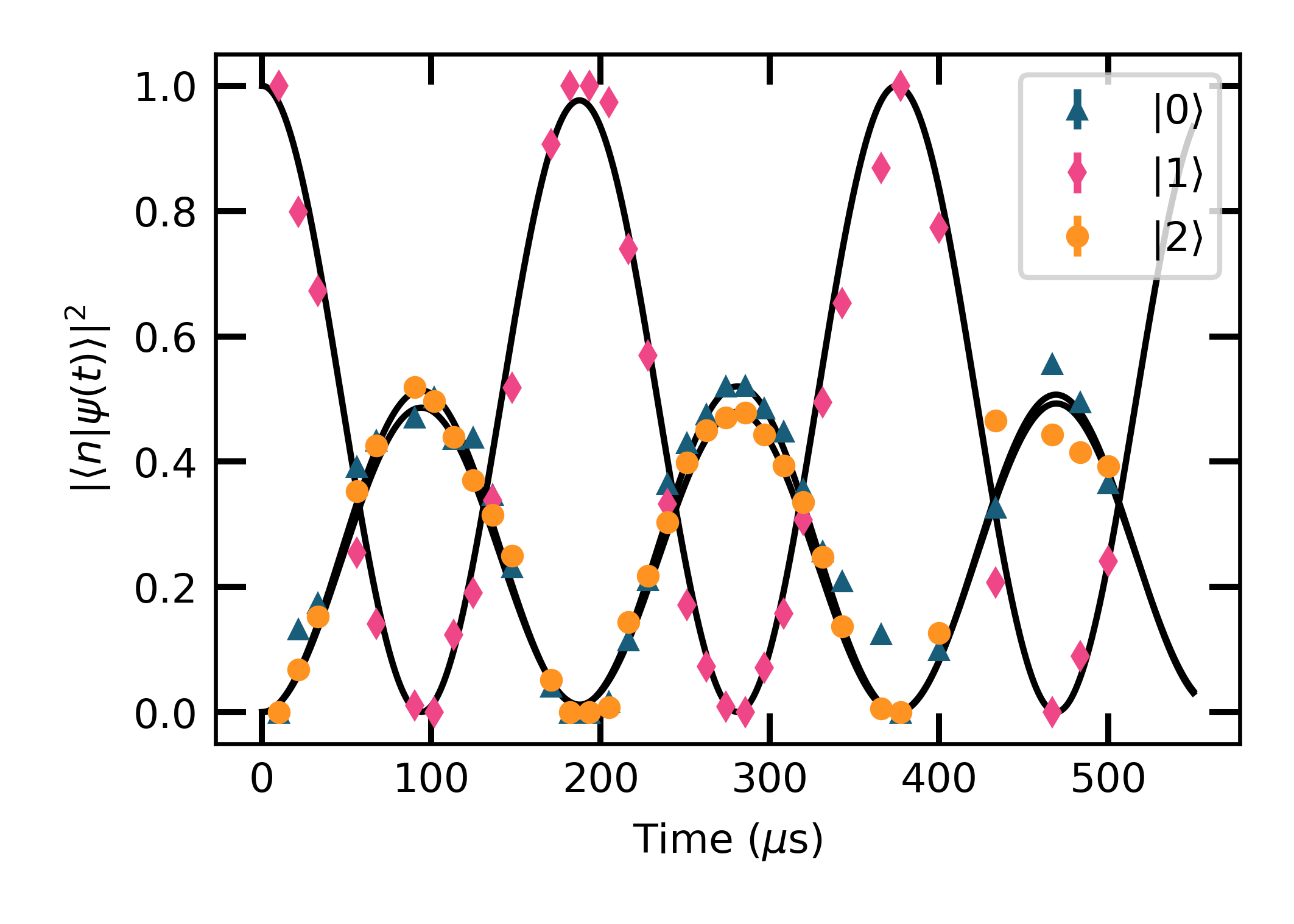}
    \caption{\textbf{3-level Rabi flopping}. A sample of experimentally measured 3-state Rabi cycles for the purpose of AWG calibration. }
    \label{fig:S3}
\end{figure}

The calibration of Rabi frequencies $\Omega_{1,2}$ is nontrivial, due to  nonlinearities in the microwave generation, amplification, and switching components; these nonlinearities mean that when both tones are applied together, each amplitude depends on the other. We account for this by constructing a cubically interpolated lookup table~\cite{renka1984triangle} that maps RF amplitudes $(A,B)$ inputted to the AWG to the resulting Rabi frequencies $(\Omega_1,\Omega_2)$. Each lookup point is determined experimentally by fitting to 3-level Rabi flopping. Figure~\ref{fig:S3} shows an example Rabi cycle.

Detuning calibration is more straightforward, but its precision is limited by ambient magnetic field drifts {and fluctuations which induce undesired detuning}. For $^{87}$Rb, the linear Zeeman shift is $\pm0.7$~MHz/G for $f=2$ and $f=1$, respectively.

\begin{figure}
    \centering
    \includegraphics[]{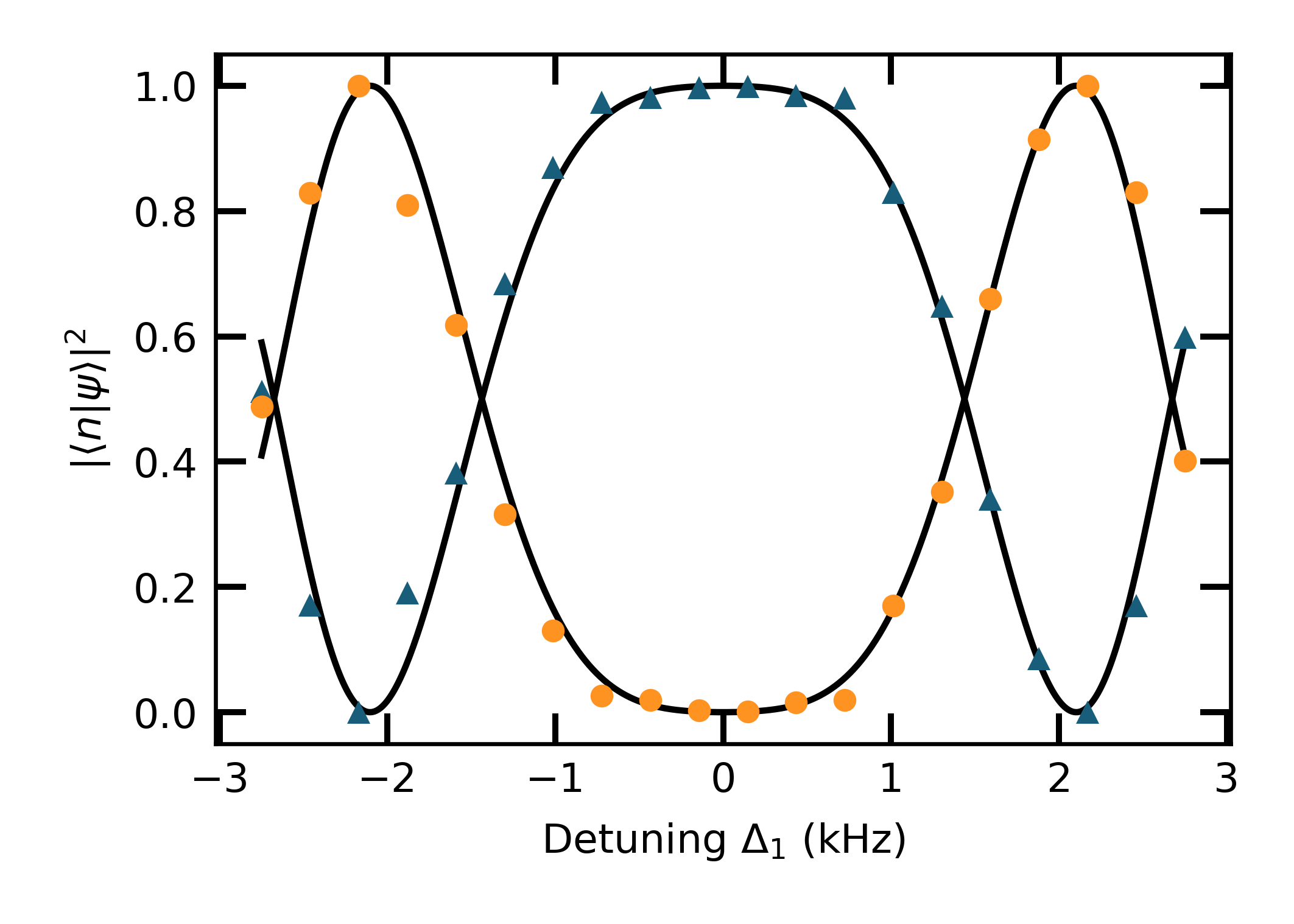}
    \caption{\textbf{Resonance calibration}. A sample experimental scan of microwave frequency over resonance, for $5\pi$ area pulses. }
    
    \label{fig:S4}
\end{figure}

To mitigate the deleterious effects of detuning drift, we use several approaches. The magnetic fields causing the Zeeman splitting are created via permanent magnets, so that we do not have to rely on potentially fluctuating current supplies driving electromagnets.  Moreover, as previously mentioned, we synchronize the experiment to the 60~Hz power line to minimize the effects of ambient ac magnetic noise. We also calibrate the resonant frequencies after every experimental cycle, due to apparent ever-present slow drifts in the background fields. Calibrations are performed by scanning the frequency until maximal population transfer is observed using an $n\pi$ pulse area pulse, with $n$ an odd integer. Population transfer around resonance is more sensitive to frequency for larger values of $n$, although we find we cannot calibrate better than $\sim100$ Hz due to shot-to-shot noise, therefore we find $5\pi$ pulses are sufficient.  Figure~\ref{fig:S4} shows a representative resonance scan.

The impact of unwanted detuning becomes more pronounced with longer pulse durations. For short pulses, phase accumulation from detuning is negligible, but longer pulses allow significant phase mismatch to develop, reducing fidelity. Therefore, we try to keep pulse sequences as short as possible.

\section{Sequence duration}
We note that the choice of ramp duration $T$ to reach the final angle requires striking a balance between precession and accuracy of the measurements taken at different angles along the ramp. We are technically limited to the maximum Rabi frequencies we can achieve, therefore ramps which are too short will deviate from the linear regime, while overly long ramps will amplify detuning errors. For the parameters achievable in our system, we chose $T = 2.5$ ms based on simulations that factor in normally distributed detuning drifts. In simulations with this duration $T$, we found that although  the non-adiabatic linear expansion starts to fail for values of $\alpha$ close to the phase transition (where the gap between energy eigenvalues becomes very small), at values away from $\alpha = 1$, the linear approximation holds well such that we readily resolve different topological phases even in the presence of technical noise and uncertainty.  

\section{Simulation}
The theoretical curves presented in this manuscript's figures were obtained by direct numerical integration of the time-dependent Schr\"odinger equation. Inputting realistic experimental parameters to this simulation shows the degree of agreement between the real evolution of the quantum system, and a perfect first order expansion to the nearly-adiabatic case. 

\begin{figure}
    \centering
\includegraphics{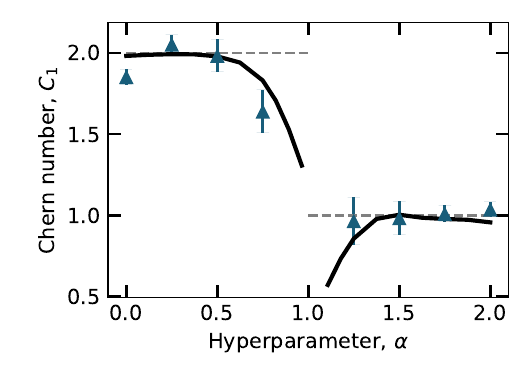}
    \caption{\textbf{Experimental simulation}. Simulation of the full experimental determination of Chern numbers, including QST, as a function of $\alpha$. Realistic laboratory parameters of Rabi frequency strengths and detuning miscalibrations are used.}
    
    \label{fig:S5}
\end{figure}

We also simulate the entirety of tomographic sequences used in the experiments presented in this work. By repeating the simulation for each experiment multiple times, followed by an additional simulation of a tomographic readout pulse, estimates of final state fidelity can be found. Furthermore, realistic detuning noise is  included in the simulation of experiments. The ramp durations of our adiabatic preparation pulses, as well as our non-adiabatic sequences to find Berry curvature, were decided upon by fully simulating our experiments from state preparation to tomography rotations, while including normally distributed detuning noise with experimentally measured variance. The results of such simulation for determining the Berry curvature via non-adiabatic evolution from north to south pole, and therefore determining the Chern number, are shown in Fig.~\ref{fig:S5}. 

\end{document}